\documentstyle[12pt]{article}

\begin{document}

\setlength{\baselineskip}{18pt}

\begin{center}
{\large \bf Weber-like interactions and energy conservation}
\end{center}

\begin{center}                             
{\bf F. Bunchaft and S. Carneiro}
\end{center}

\begin{center}
{\it Instituto de F\'{\i}sica, Universidade Federal da Bahia\\
40210-340, Salvador, BA, Brasil}
\end{center}

\begin{abstract}
Velocity dependent forces varying as
$k({\bf \hat{r}}/r)(1 - \mu \dot{r}^2 + \gamma r \ddot{r})$ (such as Weber 
force), here called {\it Weber-like} forces, are examined from the point
of view of energy conservation and it is proved that they are
conservative if and only if $\gamma=2\mu$. As
a consequence, it is shown that gravitational theories employing
Weber-like forces cannot be conservative and also yield both the
precession of the perihelion of Mercury as well as the gravitational
deflection of light.

\vspace{0.5cm}

\end{abstract}

\section{Introduction}

One and a half century ago, when Weber $[1]$ established the bases of his
electrodynamics, the energy conservation arose as a central problem of
the new theory since, for the first time, a velocity-dependent force law
was stated for a basic interaction of nature:

\begin{equation}
{\bf F}_W = \frac{q_1q_2}{4\pi \epsilon_0 r^2} {\bf \hat{r}} \left( 1 -
\frac{1}{2 c^2} \dot{r}^2 + \frac{1}{c^2} r \ddot{r} \right).
\end{equation}

\noindent Here, $q_1$, $q_2$ are the electric charges, $\epsilon_0$ is
the vacuum permittivity, $r = |{\bf r}| = |{\bf r}_1 - {\bf r}_2|$, the
separation distance from $q_2$ to $q_1$, and ${\bf \hat{r}} = {\bf r}/r$;
the dot signifies temporal derivation,
and $c$ denotes simply the ratio between the electromagnetic and the
electrostatics units of charge.

In order to face Helmholtz's criticism $[2]$, Weber introduced for the
first time a velocity-dependent potential energy

\begin{equation}
U_W = \frac{q_1q_2}{4\pi \epsilon_0 r} \left( 1 - \frac{1}{2c^2}
\dot{r}^2 \right)
\end{equation}

\noindent and succeeded to prove that ${\bf F}_W$ is derivable from $U_W$.

Some years later Tisserand $[3]$ proposed a Weber-like gravitational force law

\begin{equation}
{\bf F}_T = - \frac{Gm_1m_2}{r^2} {\bf \hat{r}} \left( 1 - \frac{1}{c^2}
\dot{r}^2 + \frac{2}{c^2} r \ddot{r} \right)
\end{equation}

\noindent derived from the Weber-like potential energy

\begin{equation}
U_T = - \frac{Gm_1m_2}{r} \left( 1 - \frac{1}{c^2} \dot{r}^2 \right),
\end{equation}

\noindent where $m_1$, $m_2$ are the gravitational masses, $G$ is the
gravitational constant, and $c$ stands also for the speed of light. With
this force, Tisserand obtained $3/8$ of the then known value for the
anomalous perihelion precession of Mercury 
and Levy $[4]$, extending this potential energy, obtained the entire
value for the precession.

In spite of its agreement with many theoretical and experimental results,
Weber electrodynamics was replaced by the Maxwell-Lorentz field theory
toward the end of the nineteenth century. And the interest in similar
forces and potentials in gravitational theories also waned. Recently
there has been a renewed interest in Weber
electrodynamics in connection with important, but still controversial,
experimental work $[5,6]$. And there has been renewed interest in
Weber-like interactions in gravitational theories, such as Assis's
Mach-like model $[7]$. With

\begin{equation}
U_A = - \frac{Gm_1m_2}{r} \left( 1 - \frac{3}{c^2} \dot{r}^2 \right)
\end{equation}

\noindent and

\begin{equation}
{\bf F}_A = - \frac{Gm_1m_2}{r^2} {\bf \hat{r}} \left( 1 - \frac{3}{c^2}
\dot{r}^2 + \frac{6}{c^2} r \ddot{r} \right)
\end{equation}

\noindent Assis reobtained the correct expression for the perihelion
precession.

More recently other theoretical Weber-like forces have been proposed to
fit gravitational observations without, however, mentioning their
conservative or nonconservative nature. Surprizingly enough, they are
indeed generally nonconservative (as shown below). This means that the
conservation of energy for Weber-like forces has not been adequately
considered.

Raguza $[8]$ extended Assis's theory by proposing the force

\begin{equation}
{\bf F}_R = - \frac{Gm_1m_2}{r^2} {\bf \hat{r}} \left( 1 - \frac{9}{c^2}
\dot{r}^2 + \frac{6}{c^2} r \ddot{r} \right),
\end{equation}

\noindent which not only yields the precession of the perihelion of
Mercury but also accounts for the gravitational deflection of light
grazing the sun.

Moreover Assis $[9]$, trying to get a unification between gravitation and
electromagnetism in a Weber-like framework, introduced a generalized
electromagnetic potential energy from which he reobtained, to second
order in $c^{-1}$, ${\bf F}_W$ (something already achieved by Phipps
$[10]$); and, to fourth order in $c^{-1}$ (from the average
electromagnetic interaction between neutral dipoles), the force

\begin{equation}
{\bf F}^*_A = - \frac{Gm_1m_2}{r^2} {\bf \hat{r}} \left( 1 -
\frac{15}{c^2} \dot{r}^2 + \frac{6}{c^2} r \ddot{r} \right)
\end{equation}

\noindent to be taken as the Weber-like gravitational force originating
from electromagnetic interaction.

Nevertheless, the conservative, or not, nature of ${\bf F}_R$ and ${\bf
F}^*_A$ has not been questioned at all. In doing so, we will not simply
return to the old Helmholtz's requirement but, instead, propose the
following inquiry: {\it What is necessary and sufficient for a Weber-like
force to be conservative? And, if so, what is the more general expression
for the potential energy?} This question, in all its generality, will be
our main concern in this short article. 

We will prove that any Weber-like force is conservative if and only if
the coefficient $\gamma$ of the acceleration term is twice the
coefficient $\mu$ of the velocity squared term (equation (\ref{1})
below). The general form that any Weber-like potential
must have is derived. It follows, in particular, that ${\bf F}_R$ and
${\bf F}^*_A$ are not conservative. A conservative Weber-like force can
involve only one adjustable parameter; so a Weber-like force cannot
simultaneously yield the gravitational deflection of light and also the
precession of the perihelion of Mercury.
This limitation does not end the matter: Generalized Weber-like forces,
to be considered elsewhere, must also be examined.

\section{The form of conservative Weber-like forces}

Let us state the following definitions.

\vspace{0.5cm}

{\it Def. 1}. A force ${\bf F}$ between two particles will be said to be
{\it Weber-like} when

\begin{equation}
\label{1}
{\bf F} = \frac{k}{r^2} {\bf \hat{r}} (1 - \mu \dot{r}^2 + \gamma r \ddot{r})
\end{equation}

\noindent where $k$ is a parameter that depends on the charges and
characterises the nature of the interaction; $\mu \ll 1$, $\gamma \ll 1$
are positive constants referred to, respectivelly, the velocity and
acceleration parameters.

\vspace{0.5cm}

Thus, a Weber-like force law has the following essential features: (i) It
is relational, in the sense that it depends only on $r$ and its time
derivatives. (ii) It is velocity and acceleration dependent, the
velocity-dependent term being of opposite sign 
to the others. (iii) It tends to a Coulomb-Newton force law when $\dot{r}
\rightarrow 0$ and $\ddot{r} \rightarrow 0$ (it reduces to such a force
law in the static case). (iv) It obeys Newton's third law in strong form.

\vspace{0.5cm}

{\it Remark 1}. Consider a system of two bodies mutually interacting
through forces ${\bf F}_1$ on body 1 and ${\bf F}_2$ on body 2 which obey
Newton's third law in strong form. The work $\delta J$ done on the bodies
during the time $dt$ by these forces and the kinetic energy $dT$ of the
system are given by

\begin{equation}
\delta J = {\bf F}_1 \cdot d{\bf r}_1 + {\bf F}_2 \cdot d{\bf r}_2 = dT =
{\bf F}_1 \cdot d({\bf r}_1 - {\bf r}_2) = {\bf F}_1 \cdot d{\bf r} = F
{\bf \hat{r}} \cdot d{\bf r} = F dr
\end{equation} 

The forces are said conservative if and only if there exists some
function $U$ of $r$ and its time derivatives, said the interaction
potential energy, such that $dU = - \delta J$, that is, such that
$d(T+U)=0$. Then

\begin{equation}
F = - \frac{1}{\dot{r}} \frac{dU}{dt}
\end{equation}

This line of reasoning can be straightforwardly extended to many-body systems.

\vspace{0.5cm}

{\it Def. 2}. A function $U(r,\dot{r})$ will be said a {\it Weber-like
potential energy} when the force derivable from $U$ is a (thus
conservative) Weber-like force (\ref{1}).

\vspace{0.5cm}

{\it Remark 2}. Let us observe that a Weber-like $U$ cannot be a function
of higher order derivatives of $r$, since then extra terms, with
derivatives of higher order then two, would appear in the derived force
law, something which is excluded by {\it Def. 1}.

\vspace{0.5cm}

The above definitions are sufficiently large to embrace all already known
relational exact Weber-like models, as well as sufficiently appropriate
to allow the unfolding
of further investigations on generalized interactions which recover
Weber-like forces in some $c^{-1}$ order of approximation. Here we will
center our attention on Weber-like forces and Weber-like
potentials\footnote{In view of the Weber-Helmholtz controverse, may be it
would be advisable to call {\it weberian} a conservative Weber-like
force, and {\it quasi-weberian} a non-conservative one.}.

The energy conservation criterion which we are searching for will be
established by the following:

\vspace{0.5cm}

{\it Theorem}. Let ${\bf F}$ be a Weber-like force (\ref{1}).

\begin{description}

\item ($1$) If ${\bf F}$ has $\gamma = 2 \mu$, then it is conservative
and its potential energy, apart an arbitrary additive constant, is given by

\begin{equation}
\label{2}
U = \frac{k}{r} ( 1 - \mu \dot{r}^2).
\end{equation}

\item ($2$) If ${\bf F}$ is conservative, then it must have $\gamma = 2
\mu$ and its Weber-like potential, apart an arbitrary additive constant,
must be given by (\ref{2}).

\end{description}

{\it Proof}. The direct assertion is immediate since

\begin{equation}
- \frac{1}{\dot{r}} \frac{d}{dt} \left[ \frac{k}{r} ( 1 - \mu \dot{r}^2)
\right] {\bf \hat{r}} = \frac{k}{r^2} {\bf \hat{r}} (1 - \mu \dot{r}^2 +
2\mu r \ddot{r}).
\end{equation}

\noindent Let us now demonstrate the inverse. In fact, if ${\bf F}$ is
conservative and we take into account the previous {\it Remarks}, there
must exist some function $U(r, \dot{r})$ such that

\begin{equation}
\label{3'}
{\bf F} = \frac{k}{r^2} (1 + \gamma r \ddot{r} - \mu \dot{r}^2) = -
\frac{1}{\dot{r}} \frac{dU}{dt} =
- \frac{\partial U}{\partial r} - \frac{1}{\dot{r}} \frac{\partial
U}{\partial \dot{r}} \ddot{r}.
\end{equation}

\noindent That is,

\begin{equation}
\label{3}
\ddot{r} \left( - \frac{1}{\dot{r}} \frac{\partial U}{\partial \dot{r}} -
\frac{k}{r} \gamma \right) = \frac{\partial U}{\partial r} +
\frac{k}{r^2} (1 - \mu \dot{r}^2)
\end{equation}

\noindent This means that, once $U(r,\dot{r})$ is introduced in
(\ref{3}), both sides must be identical expressions in the variables
$(r,\dot{r},\ddot{r})$.

As the right-hand side of (\ref{3}) does not contain $\ddot{r}$, it
follows that both sides have to be identically constant. That is,

\begin{equation}
\label{4}
\ddot{r} \left( - \frac{1}{\dot{r}} \frac{\partial U}{\partial \dot{r}} -
\frac{k}{r} \gamma \right) = b,
\end{equation}

\begin{equation}
\label{4'}
\frac{\partial U}{\partial r} + \frac{k}{r^2} (1 - \mu \dot{r}^2) = b.
\end{equation}

Since the second factor of (\ref{4}) does not contain $\ddot{r}$, the
identity can be satisfied if and only if this factor is null, that is, if
and only if $b=0$. In this case we have, by necessity, the following
system of equations in $U$:

\begin{equation}
\label{5}
- \frac{1}{\dot{r}} \frac{\partial U}{\partial \dot{r}} - \frac{k}{r}
\gamma = 0,
\end{equation}

\begin{equation}
\label{5'}
\frac{\partial U}{\partial r} + \frac{k}{r^2} (1 - \mu \dot{r}^2) = 0.
\end{equation}

Equation (\ref{5'}) leads necessarily to

\begin{equation}
\label{6}
U = \frac{k}{r} (1 - \mu \dot{r}^2) + \phi(\dot{r}),
\end{equation}

\noindent where $\phi$ is a $C^1$-differentiable function of $\dot{r}$.
Then, substituting (\ref{6}) into (\ref{5}), one gets

\begin{equation}
\label{7}
\frac{1}{\dot{r}} \frac{d\phi}{d\dot{r}} = \frac{k}{r} (\gamma - 2\mu).
\end{equation}

As in (\ref{7}) the variables $(r,\dot{r})$ are separated, the two sides
have to be identically constant, which leads to

\begin{equation}
\label{8}
\frac{k}{r} (\gamma - 2\mu) = d,
\end{equation}

\begin{equation}
\label{8'}
\frac{1}{\dot{r}} \frac{d\phi}{d\dot{r}} = d.
\end{equation}

Since $r$ is not a constant, (\ref{8}) can be satisfied if and only if
$\gamma=2\mu$ (thus $d=0$). This proves the first part of the reciprocal
assertion. Besides, $d=0$ implies, through (\ref{8'}), that
$\phi(\dot{r})$ must be an arbitrary constant and so (\ref{6}) leads to
the remaining part of the assertion.

\vspace{0.5cm}

Thus, the theorem means that any Weber-like interaction is conservative
if and only if

\begin{equation}
{\bf F} = \frac{k}{r^2} {\bf \hat{r}} (1 - \mu \dot{r}^2 + 2 \mu r \ddot{r}),
\end{equation}

\noindent and any potential energy $U$ is Weber-like if and only if it
has the form (\ref{2}).

\vspace{0.5cm}

{\it Corollary 1}. In any conservative Weber-like force law only one
parameter can be independent.

\vspace{0.5cm}

Let us observe that, for the proof of the {\it Theorem}, it has never
been necessary to restrict to $\mu \neq 0$ and/or $\gamma \neq 0$, so it
is clear that this proof also implies the following

\vspace{0.5cm}

{\it Corollary 2}. A force

\begin{equation}
{\bf F} = \frac{k}{r^2} {\bf \hat{r}} (1 - \mu \dot{r}^2)
\end{equation}

\noindent is not conservative except for $\mu=0$.

\vspace{0.5cm}

This result is already contained in Helmholtz's mathematization of the
principle of energy conservation and was the main basis for his first,
erroneous, criticism of Weber's work $[2]$. Helmholtz was not aware that
it is sufficient to add a suitable acceleration term to make the force
conservative, something which Weber was the first to do, for his
electromagnetic force.

\section{Weber-like forces and gravitational observations}

We can now restate the results of Raguza $[8]$, in the following manner:

\begin{description}

\item (1) The observed precession of the perihelion of Mercury is
obtained not only with Assis conservative force ${\bf F}_A$ (which leads
to twice of the gravitational light deflection) but indeed with any
Weber-like force with the coefficient of the acceleration term $\gamma_A
= 6/c^2$, no matter the value of the coefficient $\mu$ of the velocity
squared term.

\item (2) The gravitational deflection of light is given when $2\gamma -
\mu = 3/c^2$. Thus, to also yield the precession of the perihelion of
Mercury, it must be $\mu = 9/c^2$, giving Raguza's force ${\bf F}_R$.

\end{description}

Now we can immediately add that ${\bf F}_R$ is non-conservative, since
for this force $\gamma \neq 2\mu$. In fact, as the conditions $\gamma =
6/c^2$, $2\gamma - \mu = 3/c^2$ and $\gamma = 2\mu$ are incompatible, we
have shown the following:

\vspace{0.5cm}

{\it Proposition}. No conservative gravitational Weber-like force can
yield simultaneously the precession of the perihelion of Mercury and the
gravitational deflection of light. 

\vspace{0.5cm}

By the way, let us observe that if we assume $\gamma = 2\mu$ and $2\gamma
- \mu = 3/c^2$, that is, energy conservation and light deflection, it
results precisely Tisserand's parameters, which lead to
$\gamma_T/\gamma_A = 1/3$ of the anomalous perihelion precession.

\section{Conclusions and conjectures}

It has been shown here that any Weber-like conservative force must have a
relationship between the two parameters $\mu$ and $\gamma$, namely $\mu =
\gamma/2$. It has been further shown that this requirement is not
compatible with the independent choices of $\mu$ and $\gamma$ necessary
to yield the precession of the perihelion of Mercury and at the same time
to yield the gravitational deflection of light.

On the other hand, if the Weber-like force were not to conserve energy,
then the two body system could not be isolated; and an interaction with
the external universe would have to be assumed of the order of $1/c^2$.
This nonconservative interaction could 
not be due to radiation: electromagnetic radiation reaction is only of
the order of $1/c^3$, and gravitational radiation reaction must also be
presumed to be of order greater than the second.

Nevertheless, some ways out of this situation can be conjectured. One of
them is to assume that Weber-like forces must be improved (in analogy to
what Phipps has done, for other reasons, with Weber's electromagnetic
force) by extension to conservative generalized forces reducible to
Weber-like forces at order $c^{-2}$. Then, the Weber-like force would
only be committed to fit low velocity tests, so that Raguzas's light
deflection relation and force could be naturally dismissed and we simply
return, for low velocities, to Assis conservative force.

With respect to Assis force ${\bf F}^*_A$, it arises as the part of order
$c^{-4}$ of a generalized conservative electromagnetic force, through a
theoretical model which contains many, independent, phenomenological
assumptions which have assured perihelion precession
($\gamma^*_A=6/c^2$), but without energy conservation for ${\bf F}^*_A$
($\mu^*_A=15/c^2$). Nevertheless, there still exists the possibility that
a critical revision could lead to changes on these assumptions suitable
to obtain both results, 
that is, ${\bf F}^*_A = {\bf F}_A$.

\section*{Acknowledgements}

We would like to thanks A.K.T. Assis for the reading of the first version
of the manuscript, S. Raguza for an interesting discussion, and J.P.
Wesley for useful suggestions.

\end{document}